\documentclass[twoside]{article}

\usepackage{PRIMEarxiv}

\usepackage[utf8]{inputenc} 
\usepackage[T1]{fontenc}    
\usepackage{hyperref}       
\usepackage{url}            
\usepackage{booktabs}       
\usepackage{amsfonts}       
\usepackage{nicefrac}       
\usepackage{microtype}      
\usepackage{lipsum}
\usepackage{fancyhdr}       
\usepackage{graphicx}       
\graphicspath{{media/}}     
\usepackage[table]{xcolor}
\usepackage{multirow}
\usepackage{makecell}
\usepackage{subcaption} 

\usepackage{listings}
\usepackage{xcolor}
\lstdefinelanguage{json}{
    basicstyle=\ttfamily,
    numbers=left,
    numberstyle=\scriptsize,
    stepnumber=1,
    numbersep=8pt,
    showstringspaces=false,
    breaklines=true,
    frame=single,
    backgroundcolor=\color{gray!10},
    stringstyle=\color{red},
    literate=
     *{0}{{{\color{blue}0}}}{1}
      {1}{{{\color{blue}1}}}{1}
      {2}{{{\color{blue}2}}}{1}
      {3}{{{\color{blue}3}}}{1}
      {4}{{{\color{blue}4}}}{1}
      {5}{{{\color{blue}5}}}{1}
      {6}{{{\color{blue}6}}}{1}
      {7}{{{\color{blue}7}}}{1}
      {8}{{{\color{blue}8}}}{1}
      {9}{{{\color{blue}9}}}{1}
      {:}{{{\color{blue}:}}}{1}
      {,}{{{\color{blue},}}}{1}
      {\{}{{{\color{black}\{}}}{1}
      {\}}{{{\color{black}\}}}}{1}
      {[}{{{\color{black}[}}}{1}
      {]}{{{\color{black}]}}}{1},
}

\title{Towards a unified user modeling language for engineering human centered AI systems}
\author{
 Aaron Conrardy \\
  Luxembourg Institute of Science and Technology\\
  University of Luxembourg\\
  Esch-sur-Alzette, Luxembourg\\
  \texttt{aaron.conrardy@list.lu} \\
   \And
 Alfredo Capozucca \\
  University of Luxembourg\\
  Esch-sur-Alzette, Luxembourg\\
  \texttt{alfredo.capozucca@uni.lu} \\
  \And
 Jordi Cabot \\
  School of Coumputing and Information\\
  Luxembourg Institute of Science and Technology\\
  University of Luxembourg\\
  Esch-sur-Alzette, Luxembourg\\
  \texttt{jordi.cabot@list.lu} \\
}

\begin{document}

\newpage
\thispagestyle{empty}

\vspace*{\fill}

\begin{center}
\Large
\textbf{This is a preprint.}\\[1em]

The peer-reviewed version is published in the proceedings of the Engineering Interactive Computer Systems (EICS 2025) Workshop proceedings, as part of the third edition of the Engineering Interactive Systems Embedding AI Technologies workshop.\\[1em]

DOI: \url{https://doi.org./10.1007/978-3-032-26051-2_7}
\end{center}

\vspace*{\fill}

\newpage

\maketitle
\begin{abstract}
In today's digital society, personalization has become a crucial aspect of software applications, significantly impacting user experience and engagement. A new wave of intelligent user interfaces, such as AI-based conversational agents, has the potential to enable such personalization beyond what other types of interfaces could offer in the past. 
Personalization requires the ability to specify a complete user profile, covering as many dimensions as possible, such as potential accessibility constraints, interaction preferences, and even hobbies.
Yet, existing solutions for user modeling mostly focus on individual aspects at a very coarse level, severely limiting the potential adaptations for personalization. 
In this sense, this paper presents a unified user modeling language, aimed to combine previous approaches, both from the modeling community and other user-centric fields, in a single proposal. This language has been implemented on top of the open source BESSER low-code platform. 
Additionally, a proof of concept leveraging user profiles modeled with our language to automatically adapt a conversational agent has also been developed. 
\end{abstract}

\section{Introduction}

Users are at the center of interactive applications \cite{humanspects}. 
Yet, each user is different and has different needs and expectations based on their specific profile (age, mood, personality, etc).
Therefore, to provide an optimal user experience, applications need to adapt to each specific profile, providing a personalized interaction. Failing to do so could even lead to digital inequalities, as certain groups of users may not able to fully take advantage of an application due to their specific limitations (e.g. accessibility issues or language skills) \cite{inequality}.

Moreover, the growth of AI has increased the possibilities of user profiling and personalizing user interactions. 
Examples such as inferring the user's fatigue level based on smartwatch data using machine learning (ML) models \cite{smartwatch} or predicting a user's behavior and attitudes based on an interview using Large Language Models (LLMs) \cite{park2024generativeagentsimulations1000} have now become possible. Furthermore, based on this profile data, AI also increases the degree of personalization options \cite{aijordi}, such as automatically tuning text based on the language skills of the user \cite{llmadapt} or performing user interface layout optimization based on user data \cite{eyeformer}.

To enable all these new opportunities, we need to be able to model such user profiles in a structured and machine-readable manner, which is known as user modeling \cite{purificato2024usermodelinguserprofiling}. 
Nevertheless, there is not yet a general user modeling language \cite{liebel_human_2024,zenodo} that could be used as the \textit{de facto} standard for the model-driven engineering (MDE) community and beyond. Conversely, there are myriad of partial and overlapping solutions that result in incomplete profiles and interoperability issues. 


This paper aims to provide such user modeling language by collecting and unifying previous proposals from the modeling community, complemented with user profiling perspectives from other fields (sociology, psychology, etc.). Additionally, to illustrate the benefits of our language for personalization purposes, we have implemented a proof of concept that automatically adapts the behavior of an LLM agent based on a given user profile. 

The rest of the paper is structured as follows: Section \ref{sota} describes the state of the art on user modeling languages and current drawbacks. Section \ref{usermodelinglanguage} presents the metamodel of the proposed unified user modeling language and discusses a possible concrete syntax. The implemented proof of concept is described in Section \ref{poc} and the corresponding tool support in Section \ref{tool}. Finally, we present future work in Section \ref{roadmap} and conclude the paper in Section \ref{conclusion}.


\section{State of the art}\label{sota}
We have conducted a systematic literature review (SLR) on user modeling in MDE \cite{zenodo}.
The SLR comprised 30 papers that proposed at least some kind of abstract formalization of a user modeling language (such as a metamodel or grammar) and investigated its integration into engineering pipelines. In what follows, we briefly discuss the most relevant results.

As stated before, there is no unified user modeling language. Instead, most works rather attempted to create a new proposal from scratch, ignoring existing results, and thus, leading to a fragmented community that proposes overlapping user dimensions while, at the same time, missing others already proposed in previous works. Additionally, there is a tendency to favor more static user aspects, while neglecting more complex dynamic ones (like mood or fatigue) and/or present the dimensions at a very high level without giving enough details on the possible values that could be set for each dimension. 
A possible explanation for these limitations is that, in most of such works, modeling the user is a secondary concern rather than a first-class citizen, missing out on potential personalization opportunities.


Out of the reviewed papers, there were two exceptions that had a goal similar to ours.
Kaklanis et al. \cite{TowardsUserModelSimulation} aimed to provide a standard of user models for the purpose of user simulation in virtual environments. They explicitly mention that their focus was on an interoperable user model to describe able-bodied individuals and people with various kinds of disabilities. Thus, the scope is rather narrow compared with ours. 
Similarly, Gaspar et al. \cite{TowardsUserModelUI} created a unified user model, this time specifically for dimensions that directly affect the user interface to be presented to the user. Again, our target language aims to be more generic and enable all types of user adaptations. Even more importantly, the dimensions defined by Gaspar et al. are not detailed and stay at a very generic level which hinders the reusability of this approach.  







\section{A unified user modeling language}\label{usermodelinglanguage}

To overcome the fragmented nature of user models and provide a more complete, while at the same time, fine-grained, representation of users, we developed a new language that integrates insights from the SLR and user dimensions from other fields. This section presents the abstract and concrete syntax of the unified user modeling language.

\subsection{Abstract syntax: The unified user metamodel}\label{usermodel}
We begin by defining the abstract syntax of the user modeling language, formalized as a metamodel and represented, as usual, using the UML class diagram notation. 
As starting point for the metamodel, we took the results from the conducted SLR \cite{zenodo} as it provided a list of user dimensions and a classification for these. We have then extended such list of dimensions, e.g. by proposing the additional category "Culture", as it is regarded as another important source of influence of the user \cite{culture,humanspects}, but was missing from the reviewed literature. In some cases, dimensions are shared by multiple categories (e.g. age fits in both "Personal Information" and "Accessibility"), yet we present them as part of the most fitting one for convenience.

\setcounter{footnote}{0}
\begin{figure}[h]
    \centering
    \includegraphics[width=1\linewidth]{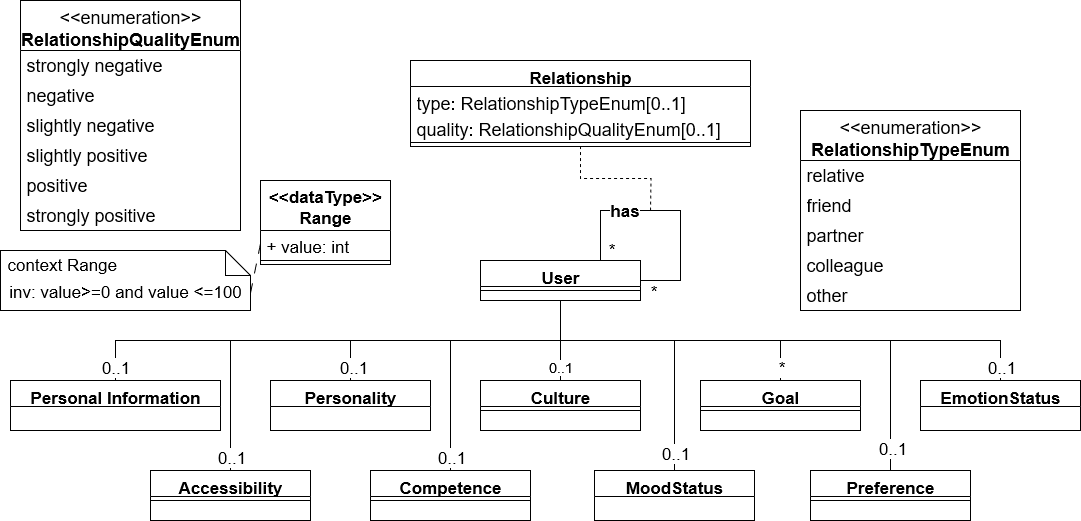}
    \caption{High-level view of unified user metamodel}
    \label{fig:highlevel}
\end{figure}

Figure \ref{fig:highlevel} depicts a high-level view of the metamodel, showcasing the main categories that characterize a user. 
Note that all categories (and the dimensions within) are optional as users may decide not to disclose that information or, in automatic profiling methods, we may not know them from the beginning. 

Furthermore, as many dimensions include a range value, we introduce a new datatype with the same name to represent integer numbers in the range of 0 to 100.
In most cases, this datatype is used when a dimension represents a score, such as the level of happiness a user is currently experiencing.
Some dimensions instead compare two opposing concepts. In those cases, the Range is used as a scale following the naming scheme "X\_To\_Y" to compare X to Y, such as "bad\_to\_good" (Figure \ref{fig:emotions}) that represents whether a user is leaning towards a bad (value leaning towards 0) or good (value leaning towards 100) mood.
Range generally offers flexibility, as depending on the requirements, one could map a predefined selection of choices to the values of Range (e.g. a binary selection of "No/Yes" could be mapped to the values 0 and 100 of Range), simulating an enumeration without altering the metamodel.

In the following, we will go over the main concepts of the metamodel, including all categories and briefly discuss their content. Due to size limitations, parts of the different metamodel are presented condensed. A complete version is available online.\footnote{\url{https://github.com/BESSER-PEARL/User-Modeling-Language}} The online version contains the complete metamodel, detailed information on the user dimensions and a mapping of reference to user dimension. 



\subsubsection{Relationship}
Information on user relationships with other users is popular in research works that deal with user models for social network platforms (e.g. \cite{P13}). However, most models only provide information on the existence of a relationship but not the type of relationship or any other relevant data.
We propose the possibility to also capture the type of relationships (relative, partner, colleague, etc.) and the quality of the relationship (Relationship class in Figure \ref{fig:highlevel}). 
This information is valuable in understanding the nuance of a relationship, as not every relationship is of positive nature and it may affect processes such as content recommendation.

\subsubsection{Personal information}
\begin{figure}[h]
    \centering
    \includegraphics[width=1\linewidth]{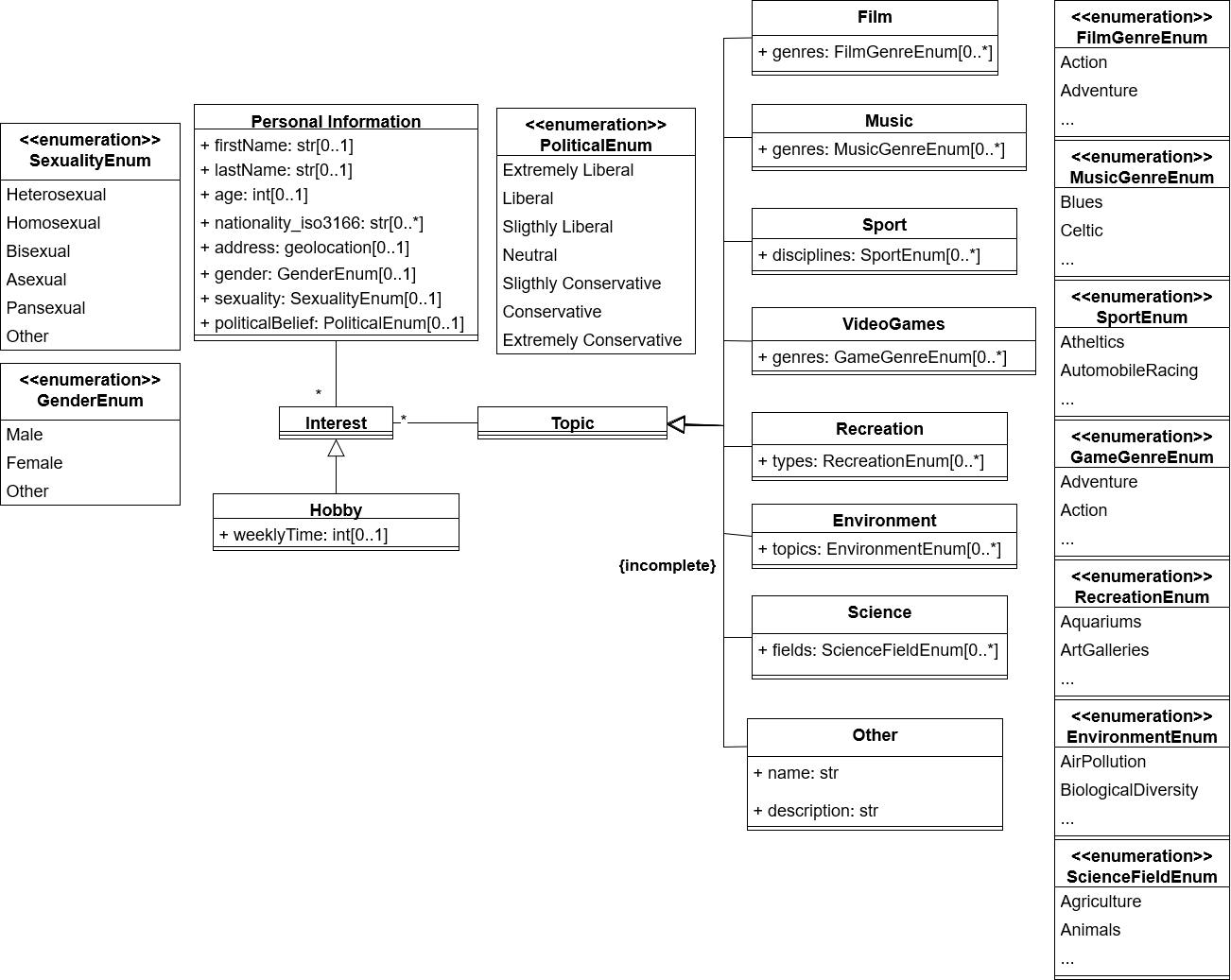}
    \caption{Personal information}
    \label{fig:personalinfo}
\end{figure}

Personal information (Figure \ref{fig:personalinfo}) describes typical information a user would enter on a profile page when registering for a new social application (e.g. a social network account).
That includes information such as the full name, age, address or gender of the user.

Regarding "Interest" and "Hobby", while both deal with engaging in topics via an activity, "Hobby" tackles a more recurrent and dedicated engagement. 
Therefore, we model "Hobby" as a stronger version of "Interest" where we indicate the number of weekly hours the user devotes to the hobby. The "Topic" class represents a non-exhaustive list of topics, derived from an ontology provided in \cite{ubiquitoususermodeling}.

\subsubsection{Competence}
\begin{figure}[h]
    \centering
    \includegraphics[width=\linewidth]{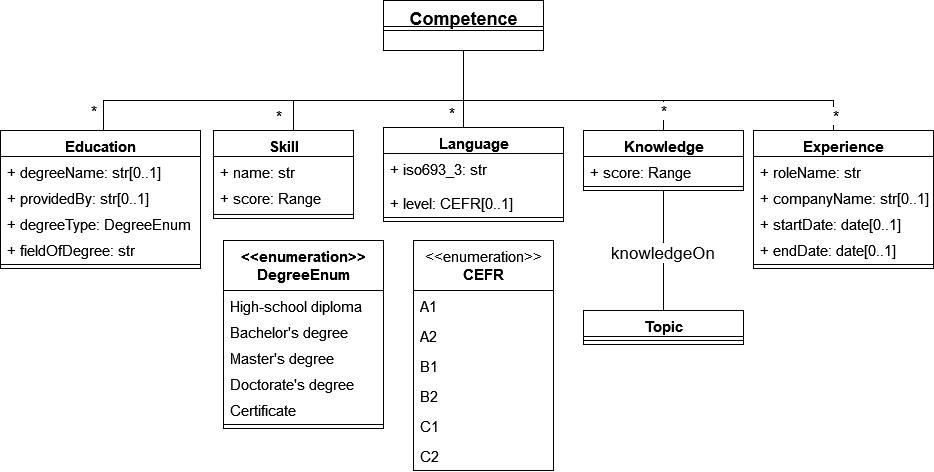}
    \caption{Competence}
    \label{fig:competences}
\end{figure}

Competence (Figure \ref{fig:competences}) describes a user's proficiency on a number of aspects such as education, language or a various number of skills and expertises. Regarding languages, we propose the usage of the Common European Framework of Reference for Languages\footnote{https://www.coe.int/en/web/common-european-framework-reference-languages/level-descriptions} (CEFR) as a standard description of a user's language skills.
Regarding "Skill" and "Knowledge", the attribute "score" is used to describe the user's ability to perform a skill or level of knowledge respectively. 
For the latter, "Knowledge" relates to the topics contained in the previously presented "Topic" class.


\subsubsection{Accessibility}
\begin{figure}[h]
    \centering
    \includegraphics[width=\linewidth]{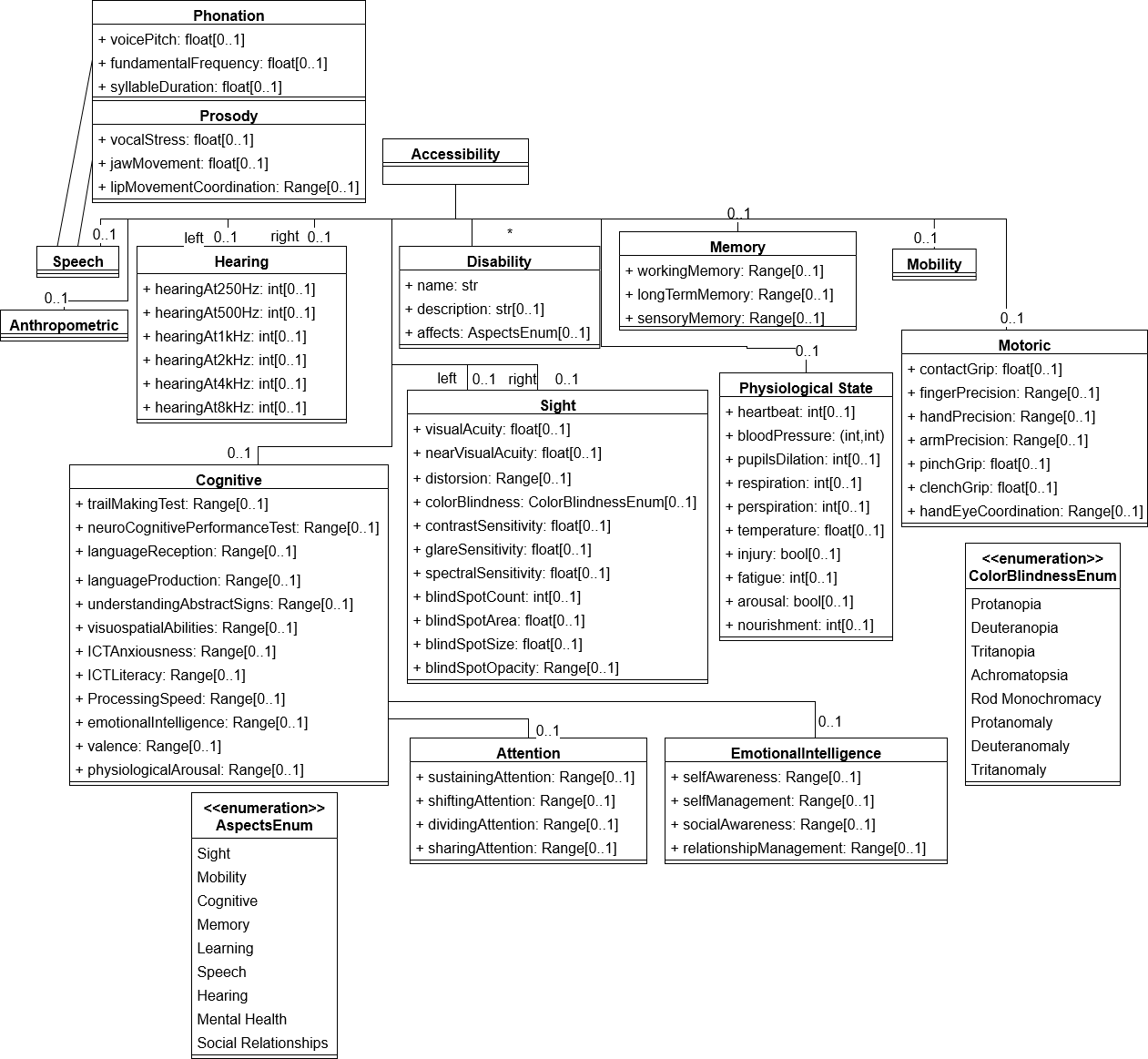}
    \caption{Accessibility}
    \label{fig:accessibility}
\end{figure}

Accessibility (Figure \ref{fig:accessibility}) refers to the accessibility needs of the user.
The "Disability" class describes a concrete disability a user might have, whereas the other classes ("Mobility", "Speech", "Hearing", etc.) describe the state of the different senses, parts of the body and body functions.
Of course, Disability and the other classes affect each other, as a completely deaf person will have the lowest hearing values.
While most of the classes are inspired  by Kaklanis et al. \cite{TowardsUserModelSimulation}, their metamodel did not contain any attribute types.
Therefore, we completed the metamodel by providing attribute types based on the used values from the results of common measurement methods for a bodily function. An example would be the usage of decibels to measure a person's hearing during an audiogram \cite{hearing}, making the choice of integers fitting.

\subsubsection{Personality}
\begin{figure}[h]
    \centering
    \includegraphics[width=\linewidth]{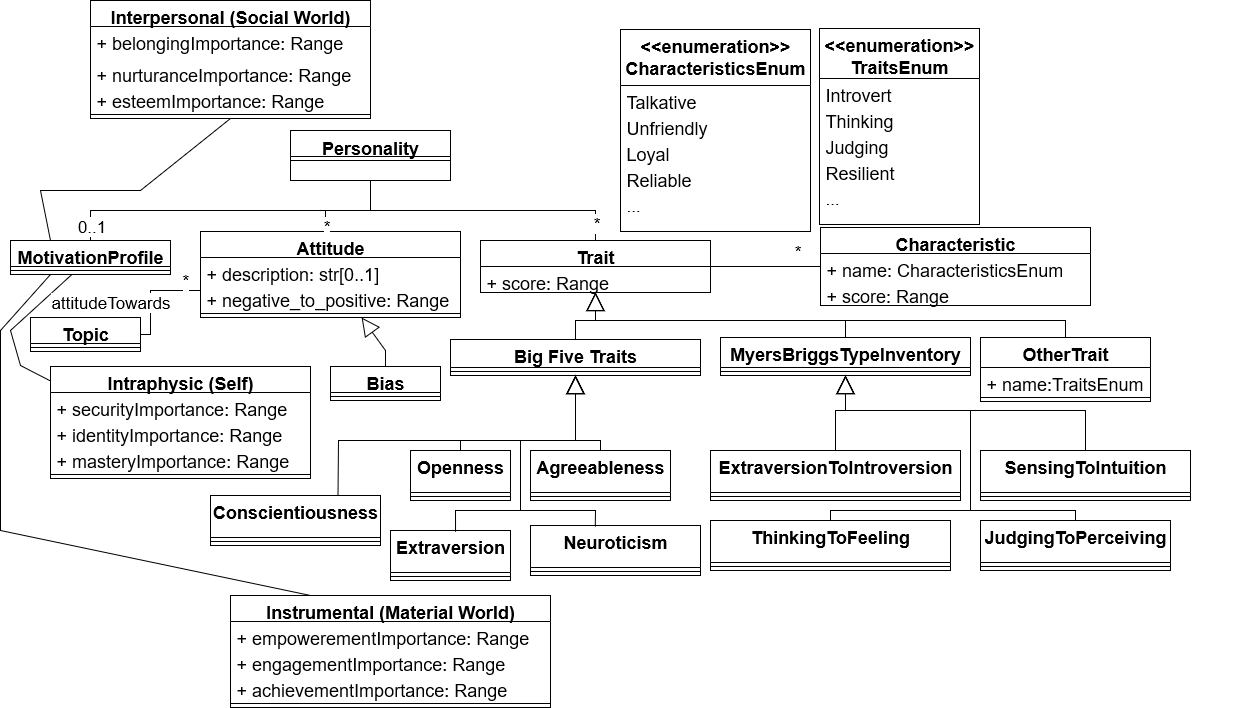}
    \caption{Personality}
    \label{fig:personality}
\end{figure}

Personality (Figure \ref{fig:personality}) tackles the user’s personality traits, that is, the stable internal characteristics of the users.
Numerous personality models were developed by psychologists to explain the nature of humans and their differences. 
For our purposes, we opted to cover personality traits from the most popular models (as proposed in \cite{ubiquitoususermodeling}), such as the Big Five personality traits \cite{bigfivebetter}.
These personality traits, represented by the "Trait" class, describe people's pattern of thoughts, feelings and behaviors \cite{cummings2019introduction}, while the "Characteristic" class describes observable and concrete feature of a person implied by a trait.
Both classes have a score to denote the extent to which a user fulfills a trait/characteristic. 

Additional dimensions affected by, or linked to personality were included, such as Motivation and Attitude.
The former is based on a unified motivational model proposed by Forbes \cite{motivation} that explains the different motivational constructs a person might lean towards to.
The latter tackles a person's opinion given a topic, be it of a negative or positive nature.
Bias in that sense is a stronger and more negative version of Attitude, often leading to an unfair treatment of a target. 

\subsubsection{Preference}
\begin{figure}[h]
    \centering
    \includegraphics[width=\linewidth]{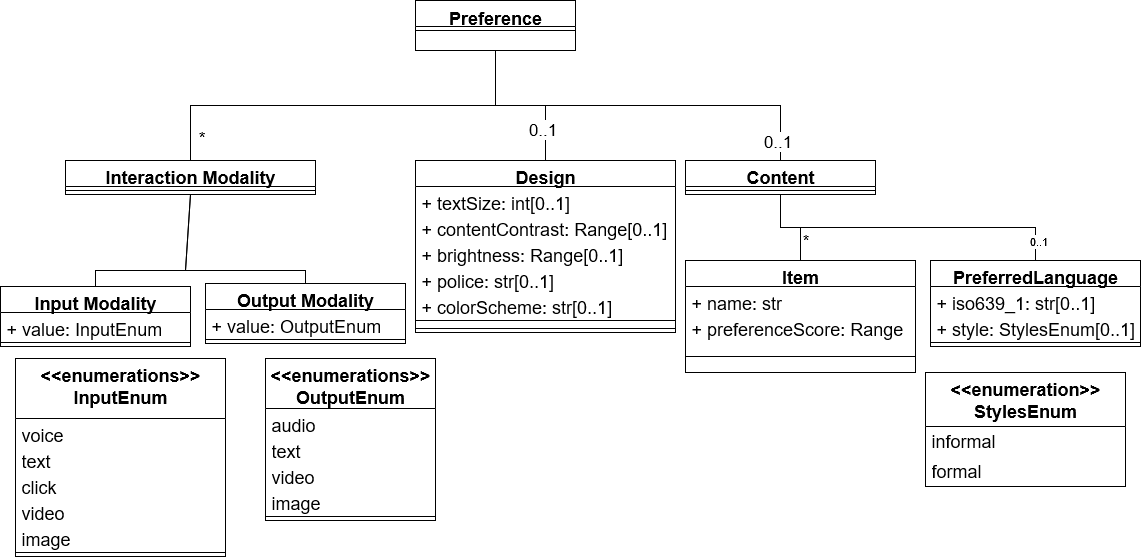}
    \caption{Preference}
    \label{fig:preferences}
\end{figure}

Preferences (Figure \ref{fig:preferences}) are used to represent a user's personal preferences, that is, a preferred option in regards to a set of choices.
In the context of a software application, the choices are narrowed to preferences of the content itself (such as preferring one topic over another or the preferred language), design preferences (such as a preferred color scheme) and the preferred interaction modality (such as audio or textual interaction).

\subsubsection{Culture}
In general, culture (Figure \ref{fig:emotions}) is linked to the social behavior, institutions and norms found in human societies and is often attributed to a specific region/location.
To model the culture of a user, we followed Hofstede's cultural dimensions \cite{hofstede}.
These consist of dimension pairs that are opposed to each other (e.g. Collectivism and Individualism), which is why we used our Range datatype.
Additionally, we added Religion as a dimension, as it is often linked to the culture of regions/locations and influences expectations during user interaction with an application \cite{religion}.


\subsubsection{Goals}
Goals (Figure \ref{fig:emotions}) focus on the objectives of a user, both while interacting with an application and outside.
Optionally, a goal can have a deadline. 
Essentially, goals provide a purpose behind the user's actions and can be both static or dynamic depending on the circumstances (a goal could appear based on an interaction that just took place).
\subsubsection{Emotions and Moods}
\begin{figure}[h]
    \centering
    \includegraphics[width=\linewidth]{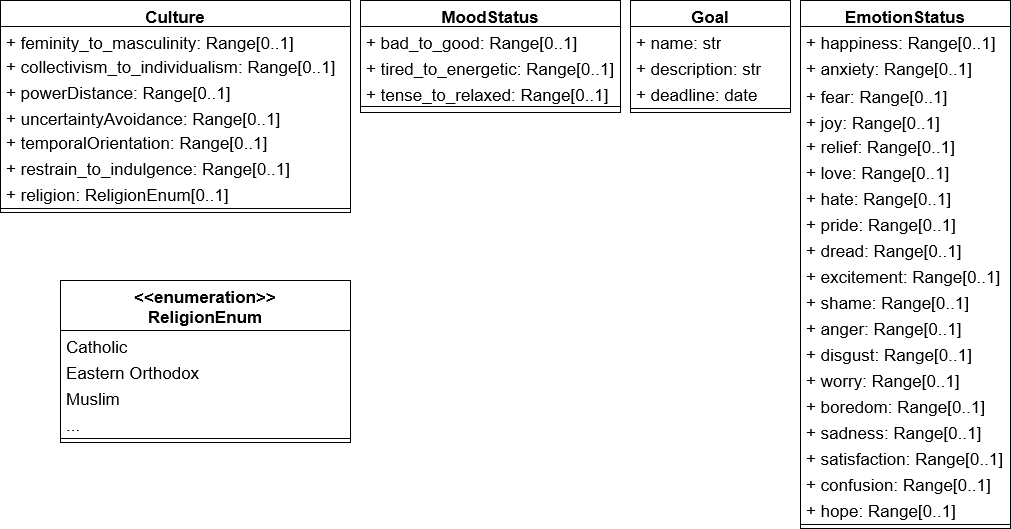}
    \caption{Culture, Mood, Goal and Emotion}
    \label{fig:emotions}
\end{figure}

Similarly to personality, there are various models in psychology that attempt to quantify emotions and moods (Figure \ref{fig:emotions}). 
Regarding the concrete difference between emotions and moods, emotions tend to be associated with a specific object or event and do not last long, while moods tend to last longer and have a less direct cause \cite{emotionvmood}.
For emotions, we decided to showcase the list of emotions as presented in GUMO \cite{gumo}. 
For moods, we opted for the three pairs of opposing moods as presented in \cite{moods}.

\subsection{Concrete syntax}\label{concretesyntax}
Based on the presented abstract syntax, we derived a textual concrete syntax to facilitate the creation of user models. In particular, we have chosen a JSON-based notation due to its popularity and its human- and machine-readable nature, which facilitates the use of user models in AI-driven personalization scenarios. Indeed, LLMs were already shown to be able to understand \cite{understandjson} as well as produce JSON \cite{mior2024largelanguagemodelsjson}.
Beyond LLMs, many tools work with JSON documents which opens the door to numerous applications exploiting user models. 

\begin{figure}[h]
    \centering
    \begin{subfigure}[c]{0.55\textwidth} 
        \centering
        \includegraphics[width=\linewidth]{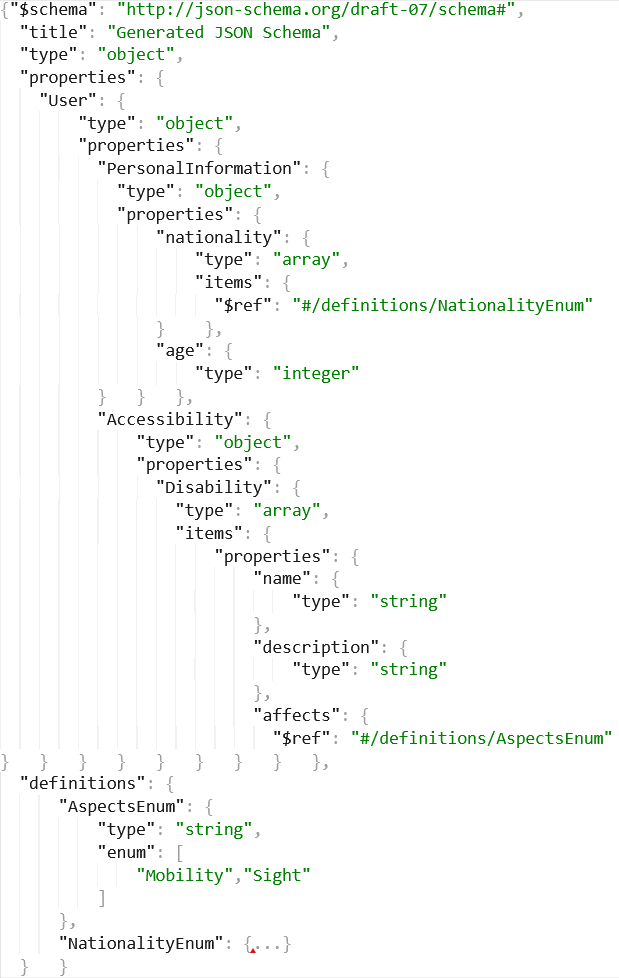} 
        \caption{JSON schema}
        \label{fig:metamodelexcerpt}
    \end{subfigure}
    \begin{subfigure}[c]{0.44\textwidth} 
        \centering
        \includegraphics[width=\linewidth]{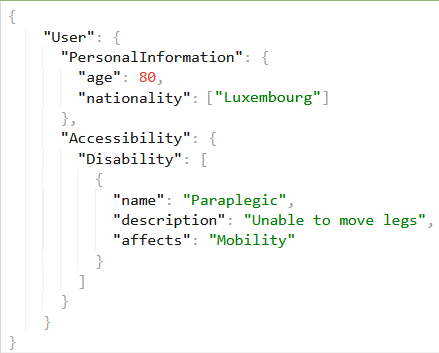} 
        \caption{Instantiation following concrete syntax}
        \label{fig:jsonsyntax}
    \end{subfigure}
    \caption{Example use of JSON syntax following the JSON schema}
    \label{fig:jsonmapping}
\end{figure}

JSON additionally allows for the definition of a JSON schema\footnote{https://json-schema.org/} that enables data consistency and validity checks of given JSON data based on predefined rules. In our case, the JSON Schema defines the concepts in the metamodel, and user models are represented as JSON objects that conform to this schema. 
Figure \ref{fig:metamodelexcerpt} shows an excerpt of the metamodel in JSON schema format whereas Figure \ref{fig:jsonsyntax} showcases the use of the concrete syntax to write user profile data that can be validated using the defined schema.

In Section \ref{tool}, we demonstrate how JSON objects representing the user model can be automatically generated from a form-based web application, further simplifying their creation.

\section{Proof of concept: Automatic conversational agent personalization}\label{poc}
To showcase the usefulness of the user modeling language, we have implemented a personalization pipeline that, given an input user model, automatically personalizes the responses of a conversational agent to increase the user experience and engagement.

\begin{figure}
    \centering
    \includegraphics[width=\linewidth]{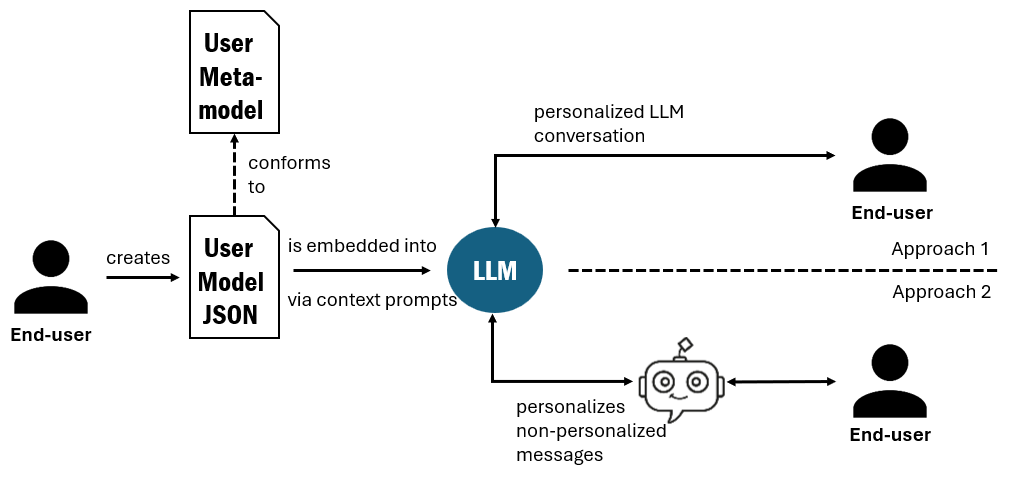}
    \caption{Implemented approaches to personalize conversational agent}
    \label{fig:approaches}
\end{figure}

Our pipeline supports two approaches to personalize conversational agents (depicted in Figure \ref{fig:approaches}), depending on the capabilities and access we have to the target LLM powering the conversational agent:
\begin{itemize}
    \item Direct personalization of LLM-agent: The information of the user model is directly embedded into the LLM via fitting context prompts. As a result, the conversation responses generated by the LLM are inherently personalized.
    \item Indirect personalization of the conversational agent: When we have an non-personalized message, whether it comes from the LLM without user context or is pre-defined by a developer, we post-process it. We do this by sending the original message and the user model to a secondary LLM, which produces a personalized version.
\end{itemize}

In the following, we describe the first approach. The process of the second approach is similar, except that it involves a separate request to an LLM with a slightly different prompt. We propose this second approach for two reasons: (1) it allows using one LLM optimized for text generation and another optimized for personalization, and (2) it supports a closed conversational agent with pre-defined conversation paths, preventing the user from engaging in open-ended dialogue.

Initially, a web application provides an interface for end-users to interact with a non-personalized LLM, with the option to upload a user model for personalization.
Once a user model that follows the JSON schema as described in Section \ref{concretesyntax} is uploaded, the following prompt context is automatically added to the conversation request "You adapt your responses based on a given user profile, such as their native language, interests, and other provided attributes. Your goal is to enhance the user's experience by tailoring your responses to their profile. This is the user's profile \{JSON file containing User Model\}".
As mentioned before, we take advantage of LLM's capabilities to understand JSON and therefore omit the transformation of the user model to a natural language form.

Once this sub-prompt is ready, it is automatically embedded into an LLM's context information.
In our case, we used OpenAI's\footnote{https://openai.com/} GPT-4o-mini model due to its current reputation as one of the most performant LLM \cite{white2024livebenchchallengingcontaminationfreellm}.
Using OpenAI's API\footnote{https://platform.openai.com/docs/overview}, one can set context information at the system level, leading to an improved upholding of defined context rules.

Finally, the web application updates its layout to indicate that the user is now interacting with a personalized LLM. 

\begin{figure}
    \centering
    \includegraphics[width=\linewidth]{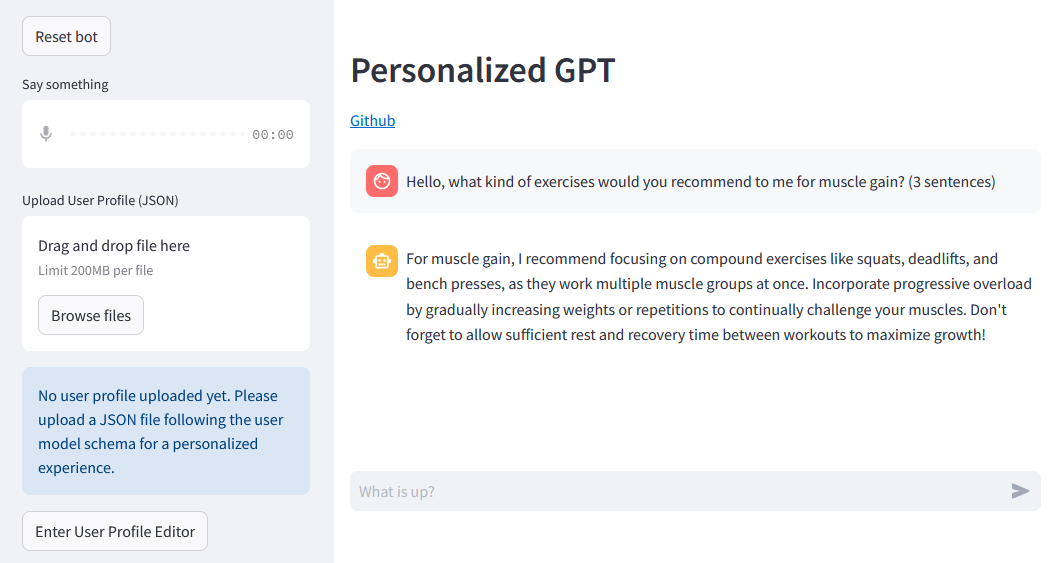}
    \caption{Overview of chat interface and example conversation with LLM (GPT-4o-mini) without personalization}
    \label{fig:frontendchat}
\end{figure}

\begin{figure}[h]
    \centering
    \begin{subfigure}{0.5\textwidth}
        \includegraphics[width=\linewidth]{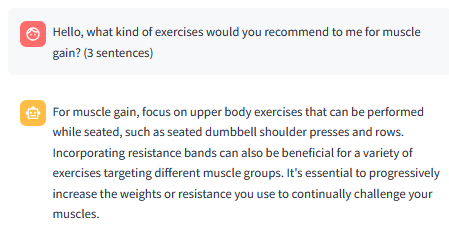} 
        \caption{Profile of paraplegic male aged 30: We notice the omission of exercises involving the lower body and an increased focus on upper body exercises}
        \label{fig:image2}
    \end{subfigure}
    \hfill
    \begin{subfigure}{0.49\textwidth}
        \includegraphics[width=\linewidth]{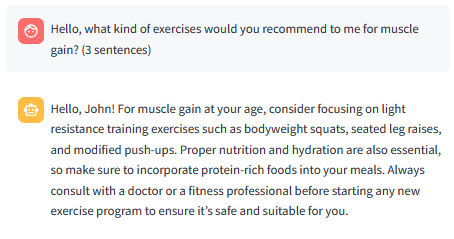} 
        \caption{Profile of male aged 80: We notice a focus on light exercises, avoiding weighted exercises and recommendations to visit a doctor to be safe due to the age}
        \label{fig:image3}
    \end{subfigure}
    \caption{Interaction with personalized LLM (GPT-4o-mini) using different user models}
    \label{fig:overall}
\end{figure}

As an example, Figure \ref{fig:frontendchat} shows the initial web application and an example interaction with GPT-4o-mini, whereas Figures \ref{fig:image2} and \ref{fig:image3} contain a comparison of the same question being asked with different user models stored in its context.
Specifically, we prompted the LLM to generate exercise recommendations for muscle gain, constrained to three sentences, reflecting the use case of an AI coach providing beginner-friendly gym advice while considering the user's profile.
We notice an adapted response based on the user's age or disability, showcasing the LLM's capabilities to adapt its responses based on the given combination of prompt and user model.
We argue that in this simple example, the personalized messages improve the user experience for the given profiles. 
Naturally, a user study would be necessary to confirm this. Such a study could consist of showing the default and personalized responses and letting participants evaluate their preferred answer.

\section{Tool support}\label{tool}
The user modeling language and the proof of concept were implemented using existing and extended components from the open-source BESSER\footnote{https://github.com/BESSER-PEARL/BESSER} low-code platform \cite{besser} and are both available online\footnote{\url{https://github.com/BESSER-PEARL/User-Modeling-Language}}.

For the user modeling language, we first used BESSER's existing modeling language called B-UML to model the abstract syntax as a structural model following its textual syntax.
We developed a new generator on top of the BESSER generators that performs a model to text transformation, transforming an input structural model into the equivalent JSON schema as defined in Section \ref{concretesyntax}.
The advantage of this generator is the possibility to easily update the abstract syntax and immediately updating the JSON schema.
This property enables future contributors to easily extend the user modeling language.

\begin{figure}
    \centering
    \includegraphics[width=0.6\linewidth]{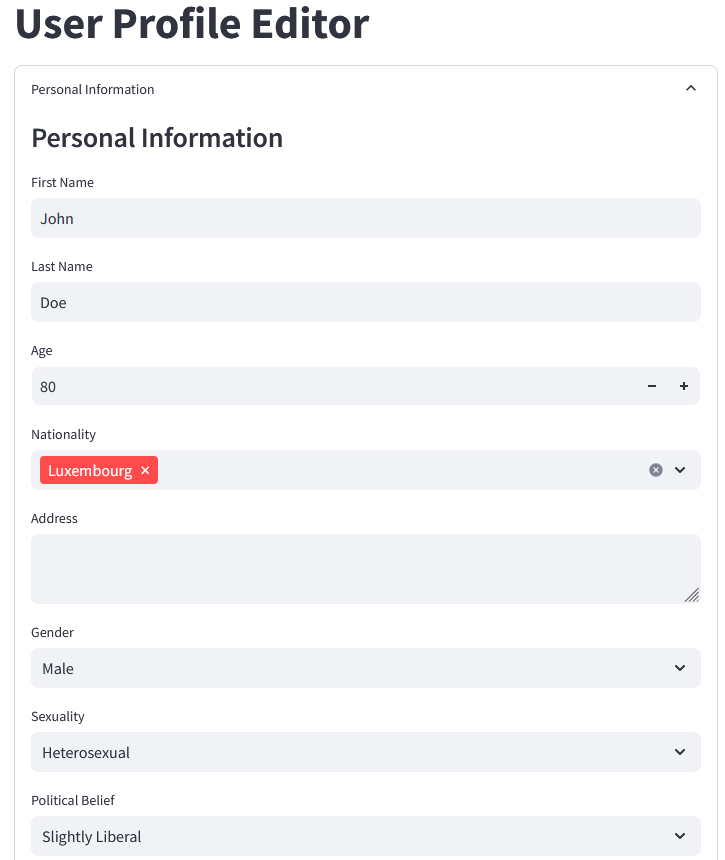}
    \caption{Forms frontend}
    \label{fig:frontend}
\end{figure}

To facilitate the usage of the concrete syntax, we developed a Streamlit\footnote{https://streamlit.io/} web-based application that lets end-users create a user model by simply filling a form, also available on the repository of the user modeling language.
Following the defined cardinalities, a user has the option to only fill out a part of the form and can leave some information out.
Figure \ref{fig:frontend} shows an excerpt of the personal information forms.
Once filled out, the information is mapped to the concrete syntax and stored as a JSON file.

The personalizable LLM was developed using BESSER's agentic framework\footnote{https://github.com/BESSER-PEARL/BESSER-Agentic-Framework}.
The framework contains wrappers for different LLMs and allows for the definition of context information that will be provided to the LLMs as high-priority context information.
Additionally, BESSER proposes the previously shown frontend (Figure \ref{fig:frontendchat}) to interact with the created personalized LLM-agents.
Both the forms and conversational agents frontend are combined, such that one can enter the forms page via the conversational agent frontend ("User Profile Editor" button in Figure \ref{fig:frontendchat}), enter the profile information and start the conversation with the personalized agent, removing the need for the end-user to manually import the user model.



\section{Future work}\label{roadmap}

\subsubsection{Interleaving with existing application modeling languages to engineer AI systems}
The proposed proof of concept does not allow to define which kind of adaptation takes place, but rather, acts non-deterministically and performs automatic adaptation outside of our control. 
The defined user modeling language should be combined with application modeling languages (e.g. IFML\footnote{https://www.ifml.org/}), specifying different kinds of behaviors based on specific values.
The resulting combined language should allow for specifying both automatic and pre-defined adaptations, taking the best of both worlds.
Existing attempts at combining user profile information and application behavior, such as \cite{yigitbas}, could act as an inspiration and be extended, avoiding re-inventing the wheel.
Correspondingly, generators shall be implemented that ingest the models and produce the adaptive application, enabling a pipeline that allows developers to both design and automatically produce (parts of) a user centered AI system.

Additionally, each language should be equipped with a graphical concrete syntax, increasing the user experience for the developers, providing straightforward drag and drop components leading to an easy and fast development.

\subsubsection{Quality evaluation of user models}
Performing quality evaluations of created user models using model-based testing, verification and validation approaches is mostly unexplored \cite{zenodo}.
Existing approaches used for application models could be adapted to be applied for user models.
Techniques to check the user model's consistency (e.g. a deaf user should not declare audio as a preferred modality) or verifying that pre-defined rules are evaluated to true (e.g. the age of a user cannot be over 200) should be included in the user modeling language.

\subsubsection{Automatic profiling of user}
Beyond static information to be provided by the end-user, we will explore how to support dynamic scenarios where applications need to adapt to users for which there is no profile information available. For such users the idea would be to discover a basic profile through the interactions with the systems to, slowly but steadily, adapt the interface to their profile.
Existing methods (e.g. emotion recognition using \cite{yigitbas} computer vision) and emerging ones (e.g. inferring the user's behaviors and attitudes based on a conversation with an LLM-agent \cite{park2024generativeagentsimulations1000}) need to be explored to allow the application to create an accurate model of the user via explicit and implicit methods. 

\subsubsection{Ethical and Privacy Considerations}
The processing of personal data is governed by legal and ethical requirements, most notably the General Data Protection Regulation (GDPR). Consequently, any personalization framework must be designed to ensure compliance with these requirements. Key aspects include obtaining and managing user consent, minimizing the collection of unnecessary data, limiting data retention to what is strictly required for processing, and pseudonymizing data when delegating processing to third parties.

\section{Conclusion}\label{conclusion}
In this paper, we have proposed a unified user modeling language that is the sum and adaptation of fragmented solutions from the MDE community and beyond.
First, the abstract syntax was formalized as a metamodel that covers a wide range of clearly defined user dimensions, allowing for a complete representation of any user and increasing the personalization potential.
Secondly, the concrete syntax is based on a JSON schema that was derived from the metamodel and enables the creation of user models that capture the users' information.
Compared with the reviewed works \cite{zenodo}, our language is much more complete, as the maximum number of categories (respectively user dimensions) covered by a solution in previous works was 5 (respectively 14), while our metamodel covers 10 categories and provides more granular details on the proposed dimensions, clearly providing a more exhaustive palette to represent users. 

Moreover, the proof of concept showcases the usefulness of the user modeling language by using the produced user models to automatically personalize a conversational agent. We argue that the provided adaptation improves the quality of the response for the given profile and, consequently, the overall user experience.

Implementing the points defined in future work will establish a pipeline to model, test and generate human centered AI systems, speeding up their development while ensuring compliance with existing regulations.


\bibliographystyle{unsrt}  
\bibliography{mybibliography}  

@article{zenodo,
  author={Conrardy, Aaron and Capozucca, Alfredo and Cabot, Jordi},
  title = {User Modeling in Model-Driven Engineering: A Systematic Literature Review},
  journal = {Journal of Object Technology},
  volume = {24},
  number = {2},
  issn = {1660-1769},
  year = {2025},
  month = may,
  editor = {},
  note = {The 21st European Conference on Modelling Foundations and Applications (ECMFA 2025)},
  pages = {2:1-14},
}

@misc{purificato2024usermodelinguserprofiling,
      title={User Modeling and User Profiling: A Comprehensive Survey}, 
      author={Erasmo Purificato and Ludovico Boratto and Ernesto William De Luca},
      year={2024},
      eprint={2402.09660},
      archivePrefix={arXiv},
      primaryClass={cs.AI}
}

@article{TowardsUserModelUI,
author = {Gaspar, Alberto and Gil, Miriam and Panach, Ignacio and Romero, Verónica},
year = {2024},
month = {01},
pages = {1-34},
title = {Towards a general user model to develop intelligent user interfaces},
volume = {83},
journal = {Multimedia Tools and Applications},
}

@article{TowardsUserModelSimulation,
author = {Kaklanis, N. and Biswas, P. and Mohamad, Y. and Gonzalez, M. F. and Peissner, M. and Langdon, P. and Tzovaras, D. and Jung, C.},
title = {Towards standardisation of user models for simulation and adaptation purposes},
year = {2016},
issue_date = {March     2016},
publisher = {Springer-Verlag},
address = {Berlin, Heidelberg},
volume = {15},
number = {1},
issn = {1615-5289},
journal = {Univers. Access Inf. Soc.},
month = mar,
pages = {21–48},
numpages = {28}
}

@inbook{culture,
author = {Plocher, Tom and Rau, Pei-Luen Patrick and Choong, Yee-Yin and Guo, Zhi},
publisher = {John Wiley \& Sons, Ltd},
isbn = {9781119636113},
title = {Cross-Cultural Design},
booktitle = {HANDBOOK OF HUMAN FACTORS AND ERGONOMICS},
chapter = {10},
pages = {252-279},
eprint = {https://onlinelibrary.wiley.com/doi/pdf/10.1002/9781119636113.ch10},
year = {2021},
}

@article{smartwatch,
  author       = {Liu, S. and Ma, C. and Chou, F. and Cheng, M. and Wang, C. and Tsai, C. and others},
  title        = {Applying a Smartwatch to Predict Work-related Fatigue for Emergency Healthcare Professionals: Machine Learning Method},
  journal      = {Western Journal of Emergency Medicine: Integrating Emergency Care with Population Health},
  year         = {2023},
  volume       = {24},
  number       = {4},
}

@inproceedings{llmadapt,
author = {Murgia, Emiliana and Pera, Maria Soledad and Landoni, Monica and Huibers, Theo},
title = {Children on ChatGPT Readability in an Educational Context: Myth or Opportunity?},
year = {2023},
isbn = {9781450398916},
publisher = {Association for Computing Machinery},
address = {New York, NY, USA},
booktitle = {Adjunct Proceedings of the 31st ACM Conference on User Modeling, Adaptation and Personalization},
pages = {311–316},
numpages = {6},
location = {Limassol, Cyprus},
series = {UMAP '23 Adjunct}
}

@inproceedings{gumo,
author = {Heckmann, Dominik and Schwartz, Tim and Brandherm, Boris and Schmitz, Michael and Wilamowitz-Moellendorff, Margeritta},
year = {2005},
month = {07},
pages = {428-432},
title = {Gumo – The General User Model Ontology},
isbn = {978-3-540-27885-6},
}

@book{moods,
author = {Lochner, Katharina and Eid, M.},
year = {2016},
month = {01},
pages = {1-455},
title = {Successful emotions: How emotions drive cognitive performance},
isbn = {978-3-658-12230-0},
}

@misc{white2024livebenchchallengingcontaminationfreellm,
      title={LiveBench: A Challenging, Contamination-Free LLM Benchmark}, 
      author={Colin White and Samuel Dooley and Manley Roberts and Arka Pal and Ben Feuer and Siddhartha Jain and Ravid Shwartz-Ziv and Neel Jain and Khalid Saifullah and Siddartha Naidu and Chinmay Hegde and Yann LeCun and Tom Goldstein and Willie Neiswanger and Micah Goldblum},
      year={2024},
      eprint={2406.19314},
      archivePrefix={arXiv},
      primaryClass={cs.CL},
}

@article{inequality,
author = {Robinson, Laura and Schulz, Jeremy and Dunn, Hopeton and Casilli, Antonio and Tubaro, Paola and Carveth, Rod and Chen, Wenhong and Wiest, Julie and Dodel, Mati and Stern, Michael and Ball, Christopher and Huang, Kuo-Ting and Blank, Grant and Ragnedda, Massimo and Ono, Hiroshi and Hogan, Bernie and Mesch, Gustavo and Kretchmer, Susan and Khilnani, Aneka},
year = {2020},
month = {07},
pages = {},
title = {Digital inequalities 3.0: Emergent inequalities in the information age},
journal = {First Monday},
}

@misc{park2024generativeagentsimulations1000,
      title={Generative Agent Simulations of 1,000 People}, 
      author={Joon Sung Park and Carolyn Q. Zou and Aaron Shaw and Benjamin Mako Hill and Carrie Cai and Meredith Ringel Morris and Robb Willer and Percy Liang and Michael S. Bernstein},
      year={2024},
      eprint={2411.10109},
      archivePrefix={arXiv},
      primaryClass={cs.AI},
}

@article{aijordi,
author = {Planas, Elena and Daniel, Gwendal and Brambilla, Marco and Cabot, Jordi},
title = {Towards a model-driven approach for multiexperience AI-based user interfaces},
year = {2021},
issue_date = {Aug 2021},
publisher = {Springer-Verlag},
address = {Berlin, Heidelberg},
volume = {20},
number = {4},
issn = {1619-1366},
journal = {Softw. Syst. Model.},
month = aug,
pages = {997–1009},
numpages = {13}
}

@inproceedings{eyeformer,
author = {Jiang, Yue and Guo, Zixin and Rezazadegan Tavakoli, Hamed and Leiva, Luis A. and Oulasvirta, Antti},
title = {EyeFormer: Predicting Personalized Scanpaths with Transformer-Guided Reinforcement Learning},
year = {2024},
isbn = {9798400706288},
publisher = {Association for Computing Machinery},
address = {New York, NY, USA},
booktitle = {Proceedings of the 37th Annual ACM Symposium on User Interface Software and Technology},
articleno = {47},
numpages = {15},
location = {Pittsburgh, PA, USA},
series = {UIST '24}
}

@article{liebel_human_2024,
	title = {Human factors in model-driven engineering: future research goals and initiatives for {MDE}},
	volume = {23},
	issn = {1619-1374},
	number = {4},
	journal = {Software and Systems Modeling},
	author = {Liebel, Grischa and Klünder, Jil and Hebig, Regina and Lazik, Christopher and Nunes, Inês and Graßl, Isabella and Steghöfer, Jan-Philipp and Exelmans, Joeri and Oertel, Julian and Marquardt, Kai and Juhnke, Katharina and Schneider, Kurt and Gren, Lucas and Happe, Lucia and Herrmann, Marc and Wyrich, Marvin and Tichy, Matthias and Goulão, Miguel and Wohlrab, Rebekka and Kalantari, Reyhaneh and Heinrich, Robert and Greiner, Sandra and Rukmono, Satrio Adi and Chakraborty, Shalini and Abrahão, Silvia and Amaral, Vasco},
	month = aug,
	year = {2024},
	pages = {801--819},
}

@book{ubiquitoususermodeling,
author = {Heckmann, Dominik},
year = {2006},
month = {01},
pages = {},
title = {Ubiquitous User Modeling},
isbn = {3898382974}
}

@misc{mior2024largelanguagemodelsjson,
      title={Large Language Models for JSON Schema Discovery}, 
      author={Michael J. Mior},
      year={2024},
      eprint={2407.03286},
      archivePrefix={arXiv},
      primaryClass={cs.DB},
}

@InProceedings{understandjson,
author="Svennberg, Kaisa
and Ekman, Jan",
title="Structuring Semi-structured Data from Building Inspection Reports Using a Large Language Model",
booktitle="Multiphysics and Multiscale Building Physics",
year="2025",
publisher="Springer Nature Singapore",
address="Singapore",
pages="508--513",
isbn="978-981-97-8313-7"
}

@InProceedings{besser,
author="Alfonso, Iv{\'a}n
and Conrardy, Aaron
and Sulejmani, Armen
and Nirumand, Atefeh
and Ul Haq, Fitash
and Gomez-Vazquez, Marcos
and Sottet, Jean-S{\'e}bastien
and Cabot, Jordi",
title="Building BESSER: An Open-Source Low-Code Platform",
booktitle="Enterprise, Business-Process and Information Systems Modeling",
year="2024",
publisher="Springer Nature Switzerland",
address="Cham",
pages="203--212",
isbn="978-3-031-61007-3"
}

@article{yigitbas,
author = {Yigitbas, Enes and Jovanovikj, Ivan and Biermeier, Kai and Sauer, Stefan and Engels, Gregor},
title = {Integrated model-driven development of self-adaptive user interfaces},
year = {2020},
issue_date = {Sep 2020},
publisher = {Springer-Verlag},
address = {Berlin, Heidelberg},
volume = {19},
number = {5},
issn = {1619-1366},
journal = {Softw. Syst. Model.},
month = sep,
pages = {1057–1081},
numpages = {25}
}

@conference{humanspects,
author={John C. Grundy.},
title={Impact of End User Human Aspects on Software Engineering},
booktitle={Proceedings of the 16th International Conference on Evaluation of Novel Approaches to Software Engineering - ENASE},
year={2021},
pages={9-20},
publisher={SciTePress},
organization={INSTICC},
isbn={978-989-758-508-1},
issn={2184-4895},
}

@article{hearing,
    author = {Patterson, Roy D. and Nimmo‐Smith, Ian and Weber, Daniel L. and Milroy, Robert},
    title = {The deterioration of hearing with age: Frequency selectivity, the critical ratio, the audiogram, and speech threshold},
    journal = {The Journal of the Acoustical Society of America},
    volume = {72},
    number = {6},
    pages = {1788-1803},
    year = {1982},
    month = {12},
    issn = {0001-4966},
    eprint = {https://pubs.aip.org/asa/jasa/article-pdf/72/6/1788/11938811/1788\_1\_online.pdf},
}

@book{cummings2019introduction,
  title     = {Introduction to Psychology},
  author    = {Cummings, Jorden A. and Sanders, Lee},
  year      = {2019},
  publisher = {University of Saskatchewan Open Press},
  address   = {Saskatoon, SK},
  isbn      = {978-0-88880-637-6},
}

@article{motivation,
author = {David L. Forbes},
title ={Toward a Unified Model of Human Motivation},

journal = {Review of General Psychology},
volume = {15},
number = {2},
pages = {85-98},
year = {2011},
eprint = { 
        https://doi.org/10.1037/a0023483
}

}

@article{hofstede,
author = {Hofstede, Geert},
year = {2007},
month = {01},
pages = {},
title = {Dimensionalizing Cultures: The Hofstede Model in Context},
volume = {2},
journal = {International Journal of Behavioral Medicine - INT J BEHAVIORAL MEDICINE},
}

@article{bigfivebetter,
title = {Looking beyond the Big Five: A selective review of alternatives to the Big Five model of personality},
journal = {Personality and Individual Differences},
volume = {169},
pages = {110002},
year = {2021},
note = {Celebrating 40th anniversary of the journal in 2020},
issn = {0191-8869},
author = {Anita Feher and Philip A. Vernon}
}

@InProceedings{P13,
author="Karam, Roula
and Fraternali, Piero
and Bozzon, Alessandro
and Galli, Luca",
editor="Abell{\'o}, Alberto
and Bellatreche, Ladjel
and Benatallah, Boualem",
title="Modeling End-Users as Contributors in Human Computation Applications",
booktitle="Model and Data Engineering",
year="2012",
publisher="Springer Berlin Heidelberg",
address="Berlin, Heidelberg",
pages="3--15",
isbn="978-3-642-33609-6"
}

@article{religion,
  title={User interface design and culture},
  author={Marcus, Aaron},
  journal={Usability and internationalization of information technology},
  volume={3},
  pages={51--78},
  year={2005},
  publisher={Erlbaum Mahwah, NJ}
}

@article{emotionvmood,
author = {Lane, Andrew and Beedie, Christopher and Terry, Peter},
year = {2005},
month = {09},
pages = {},
title = {Distinctions between Emotion and Mood},
volume = {19},
journal = {Cognition and Emotion},
}


\end{document}